%% file: ms.tex
\pdfoutput=1 

\documentclass[9pt,technote]{IEEEtran}

\usepackage[T1]{fontenc}
\usepackage[utf8]{inputenc}

\usepackage[l2tabu, orthodox]{nag} 
\usepackage[hidelinks,pdfusetitle]{hyperref} 
\usepackage[dvipsnames]{xcolor} 
\usepackage{amsmath} 
\usepackage{amsfonts}
\usepackage{amssymb}
\usepackage[autostyle=true]{csquotes}
\usepackage{placeins} 
\usepackage{graphicx}
\usepackage[font={it}]{caption}
\usepackage{lmodern} 
\usepackage{rotating} 
\usepackage{siunitx} 
\usepackage{enumitem} 
\usepackage{microtype}
\usepackage{tikz}
\usetikzlibrary{shapes,arrows,automata,positioning,chains,calc,intersections}
\usepackage{tikz-uml}
\usepackage{tikzscale} 
\usepackage{pgfplots} 
\pgfplotsset{compat=1.13}
\usepackage[sorting=none,backend=biber,style=numeric,citestyle=numeric]{biblatex} 

\graphicspath{{./Images/}}

\definecolor{IKA_Yellow}{RGB}{255,238,0}
\definecolor{IKA_Light_Gray}{RGB}{148,148,148}
\definecolor{IKA_Dark_Gray}{RGB}{92,92,92}
\definecolor{RWTH_Very_Light_Blue}{RGB}{225, 237, 249}
\definecolor{RWTH_Light_Blue}{RGB}{142,186,229}
\definecolor{RWTH_Dark_Blue}{RGB}{0,84,159}

\newcommand*\annotatedFigureBoxCustom[8]{\draw[#5,very thick,rounded corners] (#1) rectangle (#2);\node at (#4) [fill=RWTH_Light_Blue,thick,rounded corners,shape=circle,draw=RWTH_Dark_Blue,inner sep=2pt,] {\textbf{#3}};}
\newcommand*\annotatedFigureBox[4]{\annotatedFigureBoxCustom{#1}{#2}{#3}{#4}{IKA_Yellow}{white}{black}{black}}

\newenvironment {annotatedFigure}[1]{\centering\begin{tikzpicture}
	\node[anchor=south west,inner sep=0] (image) at (0,0) { #1};\begin{scope}[x={(image.south east)},y={(image.north west)}]}{\end{scope}\end{tikzpicture}}

\title{A Supervised Learning Concept for Reducing User Interaction in Passenger Cars}
\author{Marius St\"ark, Damian Backes, Christian Kehl \\Institute for Automotive Engineering (ika)\\RWTH Aachen University}

\usetikzlibrary{fadings,shapes.arrows,shadows}

\begin{document}

\maketitle

\begin{abstract}
In this article an automation system for human-machine-interfaces (HMI) for setpoint adjustment using supervised learning is presented. We use HMIs of multi-modal thermal conditioning systems in passenger cars as example for a complex setpoint selection system. The goal is the reduction of interaction complexity up to full automation. The approach is not limited to climate control applications but can be extended to other setpoint-based HMIs.
\end{abstract}


\section{Introduction}

Driver distraction is one of the leading causes of fatal and non-fatal car accidents today \cite{hundredCarNaturDrivingStudy06}\cite{klauer2014distracted}. Meanwhile, the amount and complexity of infotainment products in vehicles increases, primarily driven by customer demand \cite{johanning2015car}. Therefore, to reduce distraction, HMI complexity needs to be decreased despite increasing complexity. We will further investigate this using multi-modal thermal conditioning systems (MM-HVAC, \emph{multi-modal heating ventilation and air conditioning}) as an example.

MM-HVAC systems using thermal radiation, thermal conduction and convection can be designed to be highly energy-efficient, which is especially important for increasing the range of electric cars \cite{kuehlWaermePluginHybrid15}.

Conventional heating systems require large amounts of energy and reduce the maximum travel range of electrified vehicles. Vehicles equipped with combustion engines take this energy from the waste heat of the engine, which is impossible for electric vehicles due to their efficiency of converting electric to kinetic energy and the lower system temperature range. In electric vehicles, heat pumps and thermal radiators are required \cite{yuksel2015effects}\cite{innovativeAcousticThermalComfortATZ}. An additional benefit of non-convection-based climate systems is also the reduction of noise as shown in \cite{innovativeAcousticThermalComfortATZ} and \cite{textileApplicationsForACInVehicleInterior}.

\begin{figure}[htbp]
	\centering
	\resizebox{\linewidth}{!}{
	\begin{annotatedFigure}
		{\includegraphics[width=\linewidth]{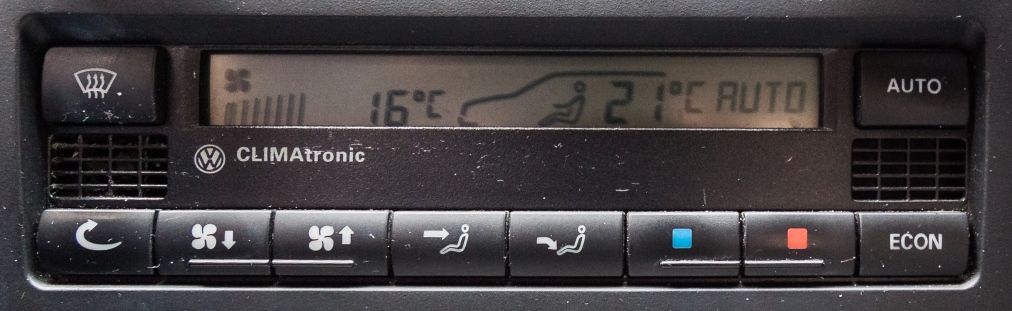}}
		\annotatedFigureBox{0.15,0.10}{0.38,0.30}{A}{0.15,0.10}
		\annotatedFigureBox{0.62,0.10}{0.851,0.3}{B}{0.62,0.10}
	\end{annotatedFigure}}
	\caption{The \enquote{CLIMAtronic} HVAC HMI for a 1999 Volkswagen Golf IV., (A) Ventilation Setpoint, (B) Cabin Temperature Setpoint.}
	\label{fig:traditional_HVAC_HMI_1}
\end{figure}

\begin{figure}[htbp]
	\centering
	\resizebox{\linewidth}{!}{
	\begin{annotatedFigure}
		{\includegraphics[width=\linewidth]{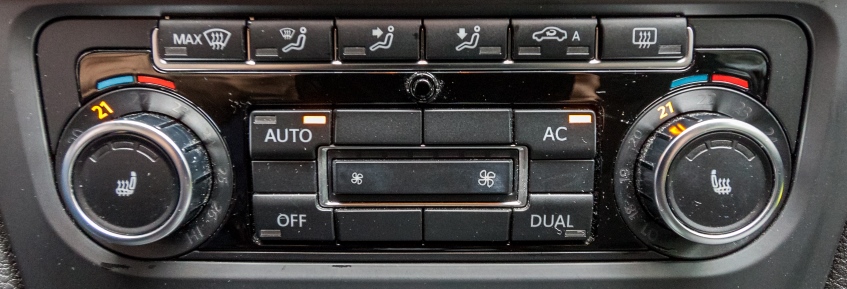}}
		\annotatedFigureBox{0.07,0.14}{0.28,0.70}{A}{0.070,0.14}
		\annotatedFigureBox{0.72,0.14}{0.94,0.70}{B}{0.72,0.14}
		\annotatedFigureBox{0.382,0.27}{0.62,0.5}{C}{0.382,0.27}
	\end{annotatedFigure}}
	\caption{The \enquote{CLIMAtronic} HVAC HMI for a 2008-2012 Golf VI., (A) Cabin and Seat Temperature Setpoint Driver, (B) Cabin and Seat Temperature Setpoint Passenger, (C) Ventilation Setpoint.}
	\label{fig:traditional_HVAC_HMI_2}
\end{figure}

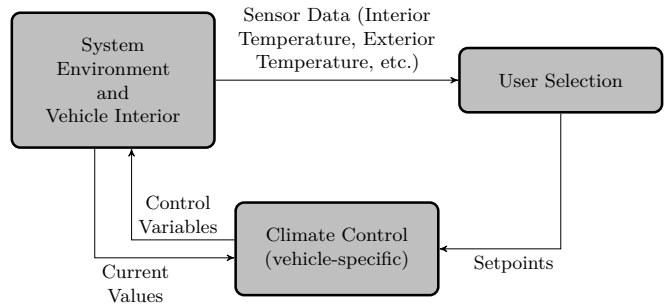
\begin{figure}[htbp]
  \centering
  \resizebox{\linewidth}{!}{\input{./Images/currentClimateControlSystems.tikz}}
  \caption{Data flow of conventional climate control systems used in current production vehicles. The user adjusts setpoints for certain control parameters. The control unit then drives the thermal actuators to reach the desired state.}
  \label{fig:currentClimateControlSystems}
\end{figure}

Typical HVAC HMIs for convection-based systems have increase/decrease-buttons or knobs for setting system setpoints like target temperature [\autoref{fig:traditional_HVAC_HMI_1}, \autoref{fig:traditional_HVAC_HMI_2}, \autoref{fig:currentClimateControlSystems}]. While MM-HVAC systems offer great ways for improving conditioning efficiency and thermal comfort, they have a lot of possible setpoints (e.g. one for every individual radiator and heat panel) and this inherent complexity requires more driver interaction compared to convection-based systems. The result of interfacing a MM-HVAC system with a typical setpoint-centric HMI is the driver being unable to take full advantage of the ability of the system to increase comfort and efficiency but simultaneously increasing distraction.

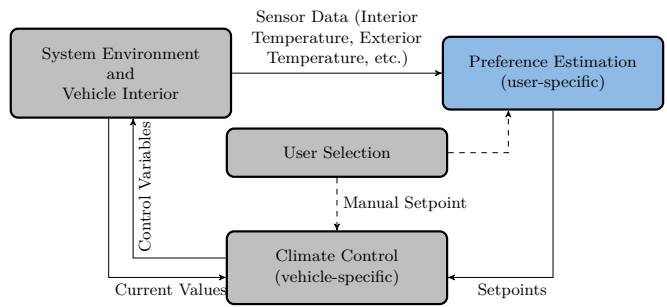
\begin{figure}[htbp]
 \centering
 \resizebox{\linewidth}{!}{\input{./Images/conceptClimateControlUserModelOnly.tikz}}
 \caption{Self-Learning climate control concept. The system from \autoref{fig:currentClimateControlSystems} is here augmented by a machine-learning-based user preference estimation unit.}
 \label{fig:conceptMachineLearningClimateControl}
 \end{figure}

To abstract from the complexity of MM-HVAC systems we propose a self-learning automation system [\autoref{fig:conceptMachineLearningClimateControl}]. It requires profound knowledge about the driver and his/her individual perception of thermal comfort. The system acquires this knowledge during operation using supervised online learning. In the following, we will present the concept using a feed-forward neural network as internal representation of the automation system but other machine learning approaches might be reasonable as well. It can be envisioned as a user-preference estimator which automatically changes the setpoints of a standard vehicle HVAC-system. The concept will be presented with respect to MM-HVAC systems in the following chapter.

\section{General Idea and Practical Application in MM-HVAC System Automation}

We define the automation unit as having a varying number of inputs $\bar{x}$ providing environment data (e.g. \enquote{cabin temperature}, \enquote{cabin humidity}, etc.) and outputs $\bar{y}$ for every setpoint of a control unit (e.g. \enquote{target temperature} is the main setpoint for a traditional HVAC control unit). 
In the beginning, the automation unit is untrained and does not know how to derive appropriate setpoints $\bar{y}$ from $\bar{x}$, the environment data. The user adjusts them manually according to his/her preferences. In order to prevent strong setpoint deviation for an untrained network, a new user can choose from a set of pretrained networks based on common thermal perception of various user types.
The automation unit then collects momentary snapshots in intervals of the manually adjusted setpoints $\bar{y}$ and corresponding environment data $\bar{x}$ to grow its training data set. 
This growing training data set is used to train the automation unit in the background using machine learning techniques. If after training a certain loss threshold for a setpoint $\bar{y}(i)$ is reached, a proposition is made to the user to hand over control of that setpoint to the automation unit.
In an ideal scenario, after a certain amount of training, all setpoints can be automated. The system has learned the preferences of the user.

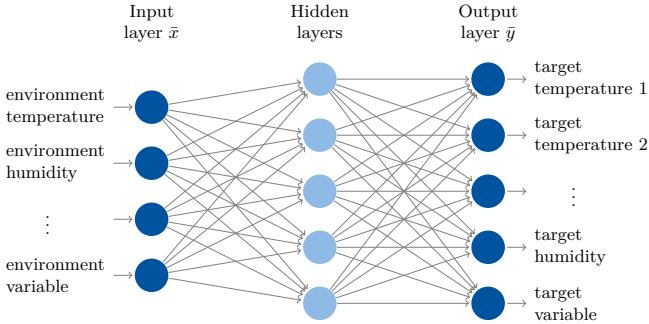
\begin{figure}[htbp]
  \centering
  \resizebox{\linewidth}{!}{\input{./Images/exampleNeuralNetUserPreferenceestimator.tikz}}
  \caption{Simplfied User Preference Estimator showing possible input and output values of the neural network}
  \label{fig:neuralNetworkUserPreferenceEstimator}
\end{figure}

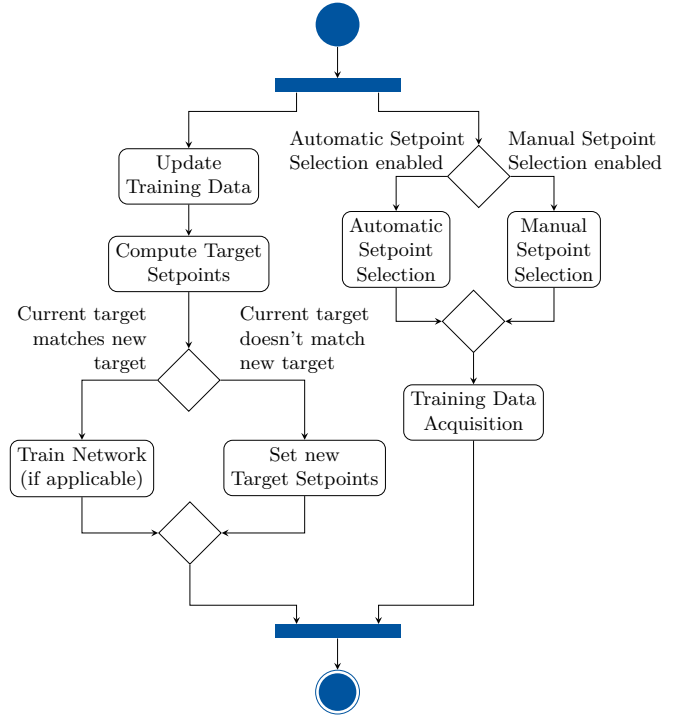
\begin{figure}[htbp]
  \centering
  \resizebox{\linewidth}{!}{\input{./Images/activity_diagramm_learnprocess.tikz}}
  \caption{Activity diagram showing the setpoint regulation, training data aquisition and neural network training}
  \label{fig:activityDiagramLearningProcess}
\end{figure}

Consider the following example: a user is generally satisfied with how the automation system works initially. However, he/she regards the average target temperature as too high. Therefore, he/she adjusts the target temperature using the HMI manually to her own preference. The system now trains the internal parameters that depend on the target temperature output (i.e. in  \autoref{fig:neuralNetworkUserPreferenceEstimator} arrows pointing directly or indirectly over multiple nodes to the target temperature node) using the user setpoint as optimal output value. After a while the system has learned that this user prefers lower temperatures. The user disables the exposure of the target temperature to the HMI, leaving its control to the automation system. \autoref{fig:activityDiagramLearningProcess} shows the different tasks of the training and automation process. 

Different machine learning methods might be possible to realize the automation concept. In the following, we will describe in more detail the necessary processes for the realization of such a system using a feed-forward neural net, taking only momentary environment values into account.

\subsection{Training Data Acquisition}

For the MM-HVAC automation, data collection is a trade-off between variety and quantity of data. Collecting many data points in rapid succession means that environment data and user settings will probably not have changed much from sample to sample, possibly leading to overfitting during training. A dynamic collection method like "acquire new sample if environment value changed by X percent" or "acquire new sample if euclidean distance of sample in parameter space to all other samples greater X" could prove to be useful. Because of the slow dynamics associated with climate control we expect average sample rates around 0.1 Hz. This means 360 data points for one hour of driving.

Some samples should be set aside for validation. The validation data set can be used to assess the quality of the automation and from this derive suggestions for the user, e.g. suggestions to change the parameters exposed to the HMI to prevent overfitting or acquire more training data for other parameters.

\begin{figure}[htbp]
  \centering
  \resizebox{\linewidth}{!}{\input{./Images/SamplingDeadTime.tikz}}
  \caption{Data Acquisition - sampling dead time is necessary to avoid generation of invalid training data due to setpoint changes by the user.}
  \label{fig:samplingDeadTime}
\end{figure}
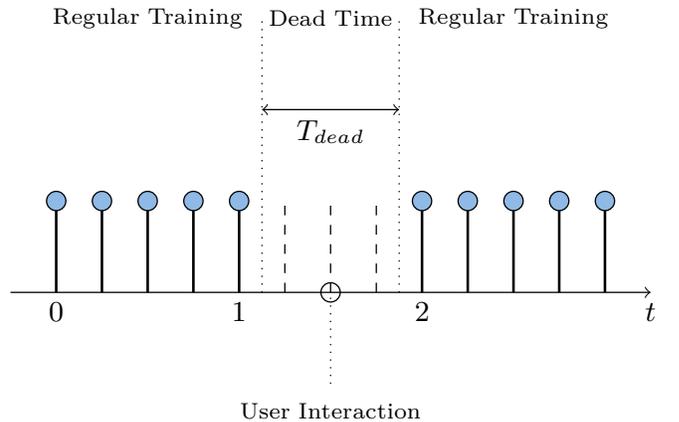

Data samples that lie chronologically next to a change of user settings should be invalidated to avoid ambiguous data, see \autoref{fig:samplingDeadTime}. We call this the sampling dead time $T_{dead}$. A sample can be added to the training data set as soon as $T_{dead} / 2$ has passed. Good values for $T_{dead}$ have to be detemined empirically.

\subsection{Training Process}

The continuously growing training data set is used to train the neural network using backpropagation. This happens during regular operation of the vehicle. Firstly, to adapt as soon as possible to changes of user preferences. Secondly, because power supply cannot be guaranteed when the car is not operated. After each training phase the resulting network is swapped with the current one. To avoid race conditions in the involved software processes, this needs to be synchronized using appropriate concurrency primitives.

\subsection{Technical Realization}

In the following, we will describe our current plans of realizing this concept as a prototype for further studies.

\begin{figure}[htbp]
	\centering
	\includegraphics[width=0.8\linewidth]{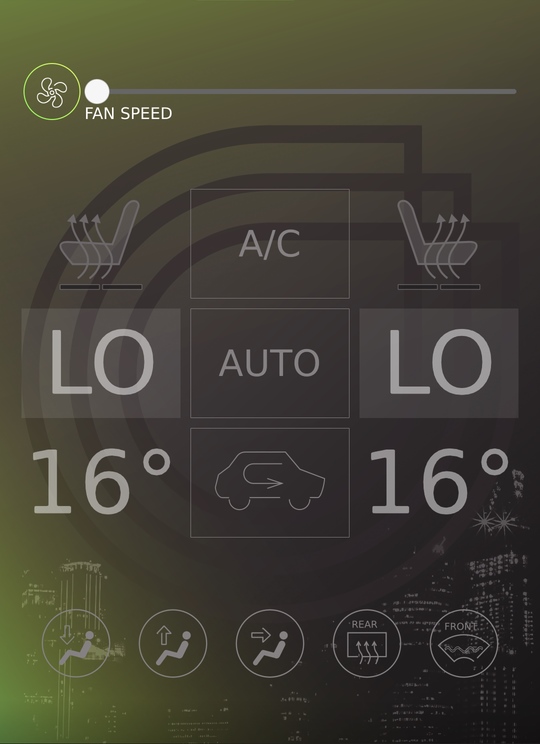}
	\caption{Automotive Grade Linux - Standard HVAC Example Application [http://docs.automotivelinux.org]}
	\label{fig:automotive_grade_linux_hvac_example}
\end{figure}

For the automation unit an additional computer running Automotive Grade Linux as operating system will be installed in a test vehicle. The computer will be connected to a touchscreen display showing the system state and providing the user interface. Since the goal of this concept is the reduction of user interface interactions during operation, a lot of effort will be related to user interface development. The HVAC sample application shipped with AGL [\autoref{fig:automotive_grade_linux_hvac_example}] serves as a design reference and will be the starting point for the user interface development.

The computer must be able to read data from the sensors and control the HVAC system using the vehicle bus (e.g. CAN), run the automation system as well as train it. The installed hardware doesn't need to be a high-end computer with GPU, FPGA or a dedicated AI processor because the automation system is assumed to be much less complex than systems for high-dimensional machine learning problems like object recognition in images and data sets are going to be relatively small. Therefore, an embedded device like a Raspberry Pi should be sufficient.

\begin{figure}[htbp]
	\centering
	\resizebox{\linewidth}{!}{\input{./Images/dataActionFlow_Automation_Unit.tikz}}
	\caption{Data Flow diagramm of the automation unit and relevant vehicle components}
	\label{fig:dataflow_Automation_Unit}
\end{figure}
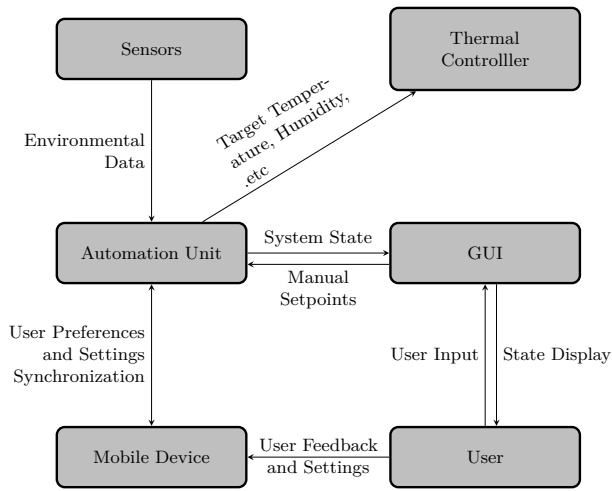

To show that the automation system can be implemented without major changes to already existing software and hardware components, standard vehicle thermal controllers, sensors and actuators are used.


After the automation unit is successfully trained the user preferences need to be stored to memory for later retrieval. If the data is stored on the automation unit itself the settings will be limited to one vehicle and the user must actively select his or her personal profile when having multiple drivers for one vehicle. Using the same user profile for multiple users will lead to unusable training data. 
To store the data on a mobile device unique to every user, e.g. a smartphone would therefore be an optimal solution. This ensures data integrity and enables portability. A user will be able switch vehicles while preserving the advantages of a user-adapted climate control system. Automatic user indentification is possible through bluetooth connection callbacks. A cloud based solution storing user specific data on a server might be sufficient as well but it would very likely compromise a cross-manufacturer usage.
If the system is used without a trained driver model, in case of a new user or in the abscence of a mobile device, a pretrained driver model stored on the automation unit is used.

The communication between vehicle computer and the smartphone storing the thermal preference data of the user is realized through a bluetooth conncection. The user can change system settings and give feedback via a smartphone app or the installed touchscreen.

\section{Summary and Prospects}

We introduced a concept to decrease HMI complexity and increase energy efficiency and thermal comfort of HVAC systems.
The intended reduction of HMI complexity is realized through an intelligent user preference estimator anticipating target setpoints of the vehicle system normally set by the user.
The user preference estimator is based on a feed-forward neural network which saves user preferences in its internal representation and is used to compute target setpoints.
The neural network is trained by manual target setpoints selected by the user in the past.
The trained networks and other settings are stored on a mobile device. Automatic bluetooth interfacing provides a seamless user experience.

While the concept is mostly relevant in the context of electric vehicles and multi-modal thermal conditioning systems using MM-HVAC systems, it is adaptable to conventional HVAC systems and cars with combustion engines as well.
Reducing driver distraction for other vehicle systems e.g. infotainment systems or cruise control, might be possible too. Adopting the concept to these domains could lead to significant increases in driving safety.

\printbibliography
\end{document}

%% file: Images/currentClimateControlSystems.tikz.tex
\tikzset{
	block/.style={
		rectangle,
		rounded corners,
		draw=black, very thick,
		minimum height=2em,
		inner sep=12pt,
		text width=2.5cm,
		text centered
	},
}

\begin{tikzpicture}[>=stealth', node distance=2 and 4]
\node[block,fill=lightgray] (environment) {System Environment \\and \\Vehicle Interior};
\node[block,fill=lightgray, right=of environment] (prefEstimator) {User Selection} ;
\node[block,fill=lightgray, below= of $(environment)!0.5!(prefEstimator)$]  (controller) {Climate Control \\(vehicle-specific)} ;

\coordinate [below=2 of controller] (belowControllerCoord);

\draw [->] (environment) -- node [above,align=center,text width=4cm] {Sensor Data (Interior Temperature, Exterior Temperature, etc.)} (prefEstimator); 
\draw [->] (prefEstimator) |- node [below left] {Setpoints} (controller); 
\draw [->] (controller.175) -| node [above right, align=center] {Control\\Variables} (environment.285); 
\draw [->] (environment.255) |- node [below right, align=center] {Current\\Values} (controller.185); 

\end{tikzpicture}

%% file: Images/conceptClimateControlUserModelOnly.tikz.tex
\tikzset{
	block/.style={
		rectangle,
		rounded corners,
		draw=black, very thick,
		minimum height=2em,
		inner sep=10pt,
		text width=3.5cm,
		text centered
	},
}

\begin{tikzpicture}[>=stealth', node distance=3 and 4]
\node[block,fill=lightgray] (environment) {System Environment \\and \\Vehicle Interior};
\node[block,,fill=RWTH_Light_Blue, right=of environment] (prefEstimator) {Preference Estimation\\ (user-specific)} ;
\node[block,fill=lightgray, below= of $(environment)!0.5!(prefEstimator)$]  (controller) {Climate Control \\(vehicle-specific)} ;
\node[block, fill=lightgray, above = 1 of controller]  (userSelection) {User Selection};

\coordinate [below= 2 of controller] (belowControllerCoord);

\draw [->] (environment) -- node [above,align=center,text width=4cm] {Sensor Data (Interior Temperature, Exterior Temperature, etc.)} (prefEstimator); 
\draw [->] (prefEstimator) |- node [below left] {Setpoints} (controller.-5); 
\draw [->] (controller.175) -| node [sloped,below right] {Control Variables} (environment.285); 
\draw [->] (environment.255) |- node [below right] {Current Values} (controller.185); 


\draw[->,dashed] (userSelection) -| (prefEstimator.220);
\draw[->,dashed] (userSelection) -- node [right] {Manual Setpoint} (controller);
\end{tikzpicture}

%% file: Images/exampleNeuralNetUserPreferenceestimator.tikz.tex
\def\layersep{3cm}
\begin{tikzpicture}[shorten >=1pt,->,draw=black!50, node distance=\layersep]
\tikzstyle{every pin edge}=[<-,shorten <=1pt]
\tikzstyle{neuron}=[circle,fill=black!25,minimum size=17pt,inner sep=0pt]
\tikzstyle{input neuron}=[neuron, fill=RWTH_Dark_Blue];
\tikzstyle{output neuron}=[neuron, fill=RWTH_Dark_Blue];
\tikzstyle{hidden neuron}=[neuron, fill=RWTH_Light_Blue];
\tikzstyle{annot} = [text width=4em, text centered]

\node[input neuron, pin={[align=left]left:environment\\temperature}] (I-1) at (0,-1) {};
\node[input neuron, pin={[align=left]left:environment\\humidity}] (I-2) at (0,-2) {};
\node[input neuron, pin=left:$\vdots \qquad \quad $] (I-3) at (0,-3) {};
\node[input neuron, pin={[align=left]left:environment\\variable}] (I-4) at (0,-4) {};

\foreach \name / \y in {1,...,5}
\path[yshift=0.5cm]
node[hidden neuron] (H-\name) at (\layersep,-\y cm) {};

\node[output neuron,pin={[pin edge={->},align=left]right:target\\temperature 1}] (O-1) at (6,-0.5) {};
\node[output neuron,pin={[pin edge={->},align=left]right:target\\temperature 2}] (O-2) at (6,-1.5) {};
\node[output neuron,pin={[pin edge={->},align=left]right:$\quad \quad \vdots$}] (O-3) at (6,-2.5) {};
\node[output neuron,pin={[pin edge={->},align=left]right:target\\humidity}] (O-4) at (6,-3.5) {};
\node[output neuron,pin={[pin edge={->},align=left]right:target\\variable}] (O-5) at (6,-4.5) {};

\foreach \source in {1,...,4}
\foreach \dest in {1,...,5}
\path (I-\source) edge (H-\dest);

\foreach \source in {1,...,5}
\foreach \output in {1,...,5}
\path (H-\source) edge (O-\output);

\node[annot,above of=H-1, node distance=1cm] (hl) {Hidden layers};
\node[annot,left of=hl] {Input layer $\bar{x}$};
\node[annot,right of=hl] {Output layer $\bar{y}$};
\end{tikzpicture}

%% file: Images/activity_diagramm_learnprocess.tikz.tex
\begin{tikzpicture}[node distance=1cm and 3cm,>=stealth]

\node[circle,fill=RWTH_Dark_Blue,minimum width=0.7cm]  (initial) {};
\node[rectangle,fill=RWTH_Dark_Blue,minimum width=2cm,minimum height=0.2cm, below=0.5cm of initial] (parallelStart1) {};
\node[diamond,draw, minimum width=1cm, minimum height=1cm, below right=1.2cm and 2cm of parallelStart1.center] (userSatisfiedDecision) {};
\node[rectangle,rounded corners,draw,minimum width=1cm, minimum height=0.8cm, left =3cm of userSatisfiedDecision,align=center] (readSetpointData) {Update\\Training Data};
\node[rectangle,rounded corners,draw,minimum width=1cm, minimum height=0.8cm, below =0.5cm of readSetpointData,align=center] (computeTargetSetpoints) {Compute Target\\Setpoints};

\node[diamond,draw, minimum width=1cm, minimum height=1cm, below = 0.9cm of computeTargetSetpoints] (targetMatchesUserSetpointDecision) {};

\node[rectangle,rounded corners,draw,minimum width=1cm, minimum height=0.8cm, below right=0.7cm and 0.3cm of targetMatchesUserSetpointDecision,align=center] (setNewTargetSetpoint) {Set new\\Target Setpoints};
\node[rectangle,rounded corners,draw,minimum width=1cm, minimum height=0.8cm, below left=0.7cm and 0.3cm of targetMatchesUserSetpointDecision,align=center] (trainNetwork) {Train Network\\(if applicable)};

\node[diamond,draw, minimum width=1cm, minimum height=1cm, below right= 0.45cm of trainNetwork] (targetMatchesUserSetpointDecisionJoin) {};

\node[rectangle,rounded corners,draw,minimum width=1cm, minimum height=0.8cm, below right=0.3cm and 0.2cm of userSatisfiedDecision,align=center] (manualSetpointSelection) {Manual\\Setpoint\\Selection};
\node[rectangle,rounded corners,draw,minimum width=1cm, minimum height=0.8cm, below left=0.3cm and 0.2cm of userSatisfiedDecision,align=center] (automaticSetpointSelection) {Automatic\\Setpoint\\Selection};
\node[diamond,draw, minimum width=1cm, minimum height=1cm, below left=0.4cm of manualSetpointSelection] (userSatisfiedDecisionJoin) {};

\node[rectangle,rounded corners,draw,minimum width=1cm, minimum height=0.8cm, below =0.5cm of userSatisfiedDecisionJoin,align=center] (trainingDataAcqui) {Training Data\\Acquisition};

\node[circle,fill=RWTH_Dark_Blue,minimum width=0.6cm, below=10cm of initial]  (finalinner) {};
\node[circle,draw=RWTH_Dark_Blue,minimum width=0.7cm] at(finalinner) (finalouter) {};

\node[rectangle,fill=RWTH_Dark_Blue,minimum width=2cm,minimum height=0.2cm, above=0.5cm of finalouter] (parallelEnd1) {};

\draw[->] (initial) -- (parallelStart1);
\draw[->] (parallelStart1.350) -- ++(0,-0.3) -| (userSatisfiedDecision);
\draw[->] (parallelStart1.190) -- ++(0,-0.3) -| (readSetpointData);
\draw[->] (readSetpointData) -- (computeTargetSetpoints);
\draw[->] (computeTargetSetpoints) -- (targetMatchesUserSetpointDecision);

\draw[->] (targetMatchesUserSetpointDecision) -| node[above,align=left] {Current target\\doesn't match\\new  target}  (setNewTargetSetpoint);
\draw[->] (targetMatchesUserSetpointDecision) -| node[above,align=right] {Current target\\matches new\\ target}  (trainNetwork);
\draw[->] (trainNetwork) |- (targetMatchesUserSetpointDecisionJoin);
\draw[->] (setNewTargetSetpoint) |- (targetMatchesUserSetpointDecisionJoin);

\draw[->] (userSatisfiedDecision) -| node[above, xshift=5mm,align=left] {Manual  Setpoint\\Selection enabled} (manualSetpointSelection);
\draw[->] (userSatisfiedDecision) -| node[above,xshift=-3mm,align=left] {Automatic  Setpoint\\Selection  enabled} (automaticSetpointSelection);
\draw[->] (manualSetpointSelection) |- (userSatisfiedDecisionJoin);
\draw[->] (automaticSetpointSelection) |- (userSatisfiedDecisionJoin);

\draw[->] (userSatisfiedDecisionJoin) -- (trainingDataAcqui);

\draw[->] (targetMatchesUserSetpointDecisionJoin) |- ( [yshift=0.3cm] parallelEnd1.170) -- ( parallelEnd1.170);
\draw[->] (trainingDataAcqui) |- ( [yshift=0.3cm] parallelEnd1.10) -- (parallelEnd1.10);

\draw[->] (parallelEnd1) -- (finalouter);
\end{tikzpicture}

%% file: Images/SamplingDeadTime.tikz.tex
\begin{tikzpicture}

\newcommand\graphheight{3}

\draw[->] (-0.5,0) -- (6.5,0) node[anchor=north] {$t$};
\draw	(0,0) node[anchor=north] {0}
		(2,0) node[anchor=north] {1}
		(4,0) node[anchor=north] {2};

\draw	(1,\graphheight) node{{\scriptsize Regular Training}};
\draw	(5,\graphheight) node{{\scriptsize Regular Training}};

\draw	(3,\graphheight) node{{\scriptsize Dead Time}};
\draw[<->] (2.25,\graphheight-1) -- (3.75,\graphheight-1);
\draw	(3,\graphheight-1) node[anchor=north] {$T_{dead}$};

\draw	(3,-1.3) node{{\scriptsize User Interaction}};
  \draw (3,0) circle (3pt);

\draw[dotted] (2.25,0) -- (2.25,\graphheight);
\draw[dotted] (3,-1) -- (3,0);
\draw[dotted] (3.75,0) -- (3.75,\graphheight);


\foreach \x in {4,...,8} {
	\draw[dashed] (\x/2,0) -- (\x/2,1);
}

\foreach \x in {0,...,4} {
  \draw[thick] (\x/2,0) -- (\x/2,1);
  \filldraw[fill=RWTH_Light_Blue] (\x/2,1) circle (3pt);
  }

\foreach \x in {8,...,12} {
  \draw[thick] (\x/2,0) -- (\x/2,1);
  \filldraw[fill=RWTH_Light_Blue] (\x/2,1) circle (3pt);
  }



\end{tikzpicture}

%% file: Images/dataActionFlow_Automation_Unit.tikz.tex
\tikzset{
	block/.style={
		rectangle,
		rounded corners,
		draw=black, very thick,
		minimum height=2em,
		inner sep=12pt,
		text width=2.5cm,
		text centered
	},
}
\begin{tikzpicture}[>=stealth, node distance = 2.5]
\node[block,fill=lightgray] at (0,0) (automationUnit) {Automation Unit};
\node[block,fill=lightgray, right=of automationUnit]  (GUI) {GUI};
\node[block,fill=lightgray, above=of automationUnit]  (sensors) {Sensors};
\node[block,fill=lightgray, right=of sensors]  (thermalController) {Thermal Controlller};
\node[block,fill=lightgray, below=of automationUnit]  (mobileDevice) {Mobile Device};
\node[block,fill=lightgray, right=of mobileDevice]  (user) {User};

\draw [->] (automationUnit) -- node [above,align=center,text width=4cm] {System State} (GUI); 
\draw [->] (automationUnit|-sensors.south) -- node [left,align=right,text width=2.5cm] {Environmental Data} (automationUnit); 
\draw [->] (automationUnit) --node [above,sloped,text width=2.5cm] {Target Temperature, Humidity, .etc} (thermalController);
\draw [<->] (mobileDevice) -- node [left,align=right,text width=2.5cm] {User Preferences and Settings Synchronization} (automationUnit);
\draw [<-] (GUI) -- node [left,align=right,text width=2.5cm] {User Input} (user);
\draw [->] ([xshift=0.2cm]GUI.south) -- node [right,align=left,text width=2.5cm] {State Display} ([xshift=0.2cm]user.north);
\draw [->] ([yshift=-0.2cm]GUI.west) -- node [below,align=center,text width=4cm] {Manual\\Setpoints} ([yshift=-0.2cm]automationUnit.east); 
\draw [->] (user) -- node [align=center,text width=2.5cm] {User Feedback and Settings} (mobileDevice);
\end{tikzpicture}